# Securing the data in cloud using Algebra Homomorphic Encryption scheme based on updated Elgamal(AHEE)

**Fahina[1], Shwetha U[2], Poorna[3], Supriya[4], Rama Moorthy H[5], Dr.Vasudeva[6]**

Dept. Computer Science & Engg. Shri Madhwa Vadiraja Institute of Technology and Management
Bantakal , Karnataka , India.

**Abstract**
*Cloud computing is the broad and diverse phenomenon. Users are allowed to store huge amount of data on cloud storage for future use. Most of the cloud service providers store data in plain text format or in secured manner but client will not be known the method in which it is stored. Homomorphic encryption is the encryption which allows the operation on cipher text thus generating an encrypted result which when decrypted, matches the result of operations performed on the plaintext. This is sometimes a desirable feature in modern communication system architectures. There are several homomorphic algorithms, one of them is AHEE. User need to use their own encryption algorithm to secure their data if required. The data needs to be decrypted whenever it is to be processed. In this paper, we have focused on providing security to the client using AHEE algorithm at the client side, for any data to be stored in the cloud.*

**Keywords—**Cloud Computing, Homomorphic Encryption, AHEE, Information Security.

## 1. INTRODUCTION

Cloud computing is the process of adopting a network of remote servers hosted on the internet to store, manage, and process data, rather than a local server or personal computer. In cloud computing, security is the main concern, because of the increment in the usage of the internet or public cloud. It is a type of Internet-based computing that provides shared computer processing resources and data to computers and other devices on interest. Cloud computing is one of the today's most attractive areas in technology. It is a model for enabling on-demand access to a shared pool of configurable computing resources (e.g., computer networks, servers, storage, applications, and services), which can be immediately provisioned and released with least management effort. Cloud computing and storage solutions grant users and enterprises with numerous facilities to store and process their data in either privately owned, or third-party data centers[3] that may be placed far from the user ranging in distance from across a city to across the world. Cloud computing entrusts on sharing of resources to attain integrity and economy of scale, comparable to a service (like the electricity grid) over an electricity network.

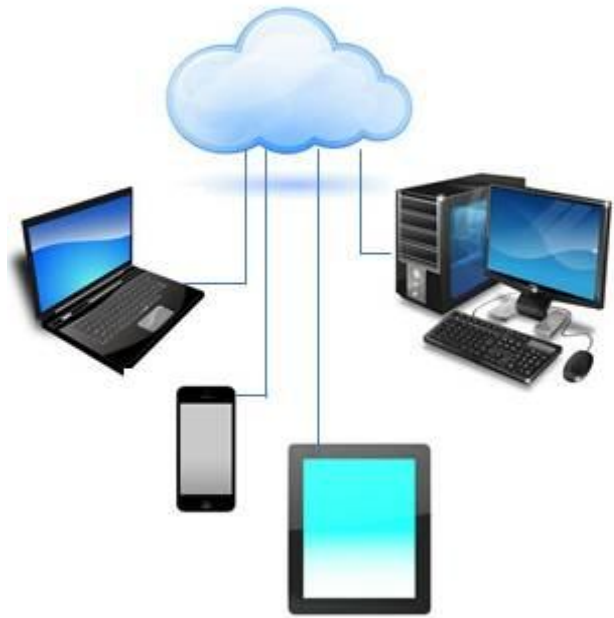

Fig 1: Cloud storage

The hardware and software part forms the cloud which is called as a public cloud where services are provided in a pay which comes under utility computing. On the other hand, private cloud grants controlled access to general public and full access to organizations and business that take advantage of such facility. Cloud computing is the sum total of SaaS and utility computing facility were data centers (small and medium) are prohibited while people can be either users or providers. Cloud computing as a utility can fascinate IT (information technology) services. Capital expenditure, as well as operation charges, can be diminished because of this innovative concept. Due to this potential capacity cloud computing is a fastest developing field in IT sector. Cloud computing and storage provides users with capabilities to store and process their data in third-party data centers. Organizations use the cloud in a variety of different service models (with acronyms such as SaaS, PaaS, and IaaS) and deployment models (private, public, hybrid, and community). When an organization nominates to store data or host applications on the public cloud, it loses its ability to have physical access to the servers hosting its information. As a result, potentially sensitive data is at risk from insider attacks. According to a recent Cloud Security Alliance Report, insider attacks are the sixth biggest threat in cloud computing. Therefore, Cloud Service Providers must ensure that thorough background checks are conducted for employees who have physical access to the servers in the data center. Additionally, data centers must be frequently monitored for suspicious activity.

The infrastructure of the cloud computing uses services and technologies, of which haven't been fully evaluated with respect to the security. One of the most critical aspects of everyday computing is the security and it is not different for the cloud computing due to importance and sensitivity of data stored on the cloud. Cloud Computing has several issues and concerns, such as regulations, data security, trust, and performances issues.

## 2. BACKGROUND THEORY

A total of security threats are correlated with cloud data services: not only long-established security threats, like network eavesdropping, illegal invasion and denial of service attacks but also precise cloud computing threats, such as side channel attacks, virtualization vulnerabilities and exploitation of cloud can also be advertised.

In the classical approaches of the algorithm public key and private key are managed by server/third party. It sends the public key to the client. Using this public key client encrypts the data and sends to the cloud where cloud decrypts the data sent by the clients and stores it in the data storage.

The data can be viewed by the server and there are chances that the data can be compromised. Several trends are opening up the era of Cloud Computing, which is an Internet-based development and use of computer technology. The ever cheaper and more powerful processors, together with the software as a service (SaaS) computing architecture, are transforming data centers into pools of computing service on a huge scale. The increasing network bandwidth and reliable yet flexible network connections make it even possible that users can now subscribe high quality services from data and software that reside solely on remote data centers.

As a result, users are at the mercy of their cloud service providers for the availability and integrity of their data. Recent downtime of Amazon's S3 is such an example. From the perspective of data security, which has always been an important aspect of quality of service, Cloud Computing inevitably poses new challenging security threats for number of reasons. Firstly, traditional cryptographic primitives for the purpose of data security protection can-not be directly adopted due to the users' loss control of data under Cloud Computing. Therefore, verification of correct data storage in the cloud must be conducted without explicit knowledge of the whole data.

The major gaps in cloud computing are availability of services, data lock-in, data confidentiality and auditability, data transfer bottlenecks, performance unpredictability, scalable storage, bugs and software licensing. Cloud computing is associated with the tradeoffs between cost and security. Security concerns associated with cloud computing fall into two broad categories: security issues faced by cloud providers (organizations providing software-, platform-, or infrastructure-as-a-service via the cloud) and security issues faced by their customers (companies or organizations who host applications or store data on the cloud).The responsibility is shared, however. The provider must ensure that their infrastructure is secure and that their clients' data and applications are protected, while the user must take measures to fortify their application and use strong passwords and authentication measures.

Moving data into the cloud offers extreme satisfaction to users since they don't have to care about the complications of direct hardware management. The pioneer of Cloud Computing vendors, Amazon Simple Storage Service (S3) and Amazon Elastic Compute Cloud (EC2) are both well-known examples. While these internet-based online services do provide enormous amounts of storage space and customizable computing resources, this computing platform shift, however, is eliminating the responsibility of local machines for data maintenance at the same time. Security is essential to serve the integrity, confidentiality, and availability of the information system resource stacks but also specified cloud computing threats, such as channel attacks, virtualization vulnerabilities.

Homomorphic encryption would allow different services to chain together without the data getting exposed to each of those services. Homomorphic encryption schemes contain property of some cryptographic algorithms. This enables their use in cloud computing environment for ensuring the confidentiality of processed data. Many other secure systems can be created by homomorphic property of various cryptosystems.

Homomorphic encryption enables use in cloud computing environment for ensuring the confidentiality of processed data. In addition the homomorphic property of various cryptosystems can be used to create many other secure systems, for example, secure voting systems,[2] collision-resistant hash functions, private information retrieval schemes, and much more.

In this paper, we have concentrated on the security of the data to be stored in the cloud.

## 3. LITERATURE REVIEW

There is 2 ways of encryption, symmetric and asymmetric encryption. The symmetric key algorithm uses the same key for encryption as well as decryption. An Asymmetric key algorithm uses different keys for encryption and decryption that is private key and public key.

Symmetric-key algorithms are algorithms for cryptography that use the same cryptographic keys for both encryptions of plaintext and decryption of ciphertext. The keys may be identical or there may be a simple transformation to go between the two keys [citation needed]. The keys, in practice, represent a shared secret between two or more parties that can be used to maintain a private information link. Message authentication codes can be constructed from symmetric ciphers. However, symmetric ciphers cannot be used for non-repudiation purposes except by involving additional parties. Another application is to build hash functions from block ciphers. One-way compression functions for descriptions of several such methods. The various types of symmetric encryption are AES, DES, Triple DES, BlowFish, TwoFish.

Asymmetric cryptography is a scheme that needs pairs of keys: public keys which may be noted universally, and private keys which are known only to the owner. This concludes two objectives: authentication, which is when the public key is used to certify that a master of the paired private key sent the message and encryption, whereby only the master of the paired private key can decrypt the message encrypted with the public key. Public key algorithms, dissimilar to symmetric key algorithms, do not feel the necessity of a secure channel for the basic exchange of one (or more) secret keys between the parties. Because of the computational complexity of asymmetric encryption, it is commonly used only for small blocks of data, occasionally the transfer of a symmetric encryption key (e.g. a session key). This symmetric key is then used to encrypt the rest of the possibly lengthy message sequence. The symmetric encryption/decryption is established on uncomplicated algorithms and is much rapid. Some of the asymmetric encryption are Diffie-Hellman, RSA, ElGamal.

The concept of Homomorphic encryption was suggested for the first time by Ronald Rivest, Leonard Adleman, and Michael Dertouzos[4]. Since then, little progress has been made for 30 years. The encryption system of Shafi Goldwasser and Silvio Micali was proposed in 1982 was a provable security encryption scheme which reached a remarkable level of safety, it was an additive Homomorphic encryption, but it can encrypt only a single bit. In the same concept, Pascal Paillier has also proposed a provable security encryption system that was also an additive Homomorphic encryption. A few years later, Dan Boneh, Eu-Jin Goh and Kobi Nissim [3] invented a system of provable security encryption, with which the unlimited number of additions but only one multiplication can be performed. Homomorphic encryption is the encryption scheme where the operations are performed on the encrypted data rather than on the original data with providing the result as it is done on the plain text. The complex mathematical operations can be performed on the ciphertext without changing the nature of the encryption [4].

Homomorphic encryption would allow the chaining together of different services without exposing the data to each of those services. For example, a chain of different services from different companies could calculate 1) the tax 2) the currency exchange rate 3) shipping, on a transaction without exposing the unencrypted data to each of those services[1].

There are several partially homomorphic cryptosystems, and also a number of fully homomorphic cryptosystems. Although a cryptosystem which is unintentionally malleable can be subject to attacks on this basis, if treated carefully homomorphism can also be used to perform computations securely.

The different homomorphic encryption schemes are- Algebra Homomorphic encryption scheme based on updated ElGamal (AHEE), Non-interactive exponential homomorphic encryption algorithm(NEHE), Enhanced Homomorphic cryptosystem(EHC),Brakerski-Gentry-Vaikuntanathan (BGV).

**Table 1:** Comparison between various homomorphic encryption scheme [1]

|          | Add-Homo | Multi-Homo | Mixed-Homo | Applications |
|----------|----------|------------|------------|--------------|
| Paillier | ✓        | x          | x          | e-voting system, threshold scheme |
| RSA      | x        | ✓          | x          | To secure internet, Banking and credit card Transactions |
| ElGamal  | x        | ✓          | x          | In Hybrid System |
| BGV      | x        | x          | ✓          | For Security of integer polynomials |
| EHC      | x        | x          | ✓          | Efficient Secure Message Transmission in Mobile AdHoc Networks |
| NEHE     | x        | x          | ✓          | Active networks, e-commerce Based on mobile agent, computing grid |
| AHEE     | x        | x          | ✓          | Secure multi party computation, electronic voting and moblie cipher |

In table 1, we have shown the comparison of various algorithms. BGV (Brakerski-Gentry-Vaikuntanathan) is an asymmetric encryption which is based on the encryption of bits. EHC (Enhanced Homomorphic cryptosystem) is the type of homomorphic encryption with its numerous application based on real time. NEHE (Non-interactive exponential homomorphic encryption algorithm) is the type of encryption based on exponential functions and polynomial functions. The AHEE algorithm has the application: secure multi-party computation, electronic voting, and mobile cipher. Since our project is to secure the data in cloud AHEE is algorithm is suitable.

| |
|---|
| **Step 1**: Select any 2 prime numbers say **p** and **q** |
| **Step 2**: Calculate the product of those 2 prime numbers ie, **N=p*q** <br> where p and q being confidential and N is public |
| **Step 3**: Select random number **x** and a **root g of GF(p)** <br> Where both x and g are smaller than p |
| **Step 4**: Calculate **y=$g^x$ mod p** Use this y for encryption |
| **Step 5**: Encryption will be performed in following 2 steps: <br> 1. Select the random integer number **r** and apply following homomorphic encryption <br> $E_1(M)=(M+r*p)$ mod N <br> 2. Select random integer number **k** and the encryption algorithms are: <br> $E_g(M)=(a,b)=(g^k$ mod p , $y^k E_1(m)$ mod p) |
| **Step 6**: Decrypted algorithm **$D_g()$** is <br> $M=b*(a^x)^{-1}$ **(mod p)** |

**Fig 2:** AHEE homomorphic scheme[1]

The security in AHEE is the highest level of security. This algorithm is proved to be secure because it is able to prevent from plaintext attack because of the random number k in the E1().

## 4. PROPOSED METHODOLOGY

By utilizing Algebra Homomorphic Encryption Scheme Based On Updated ElGamal (AHEE), we achieve data correctness and security in storage.

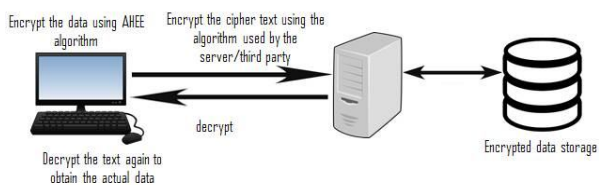

**Fig 3:** encrypting the data using AHEE and storing in cloud

AHEE is the modified form of the digital signature standard DSS presented by the NIST in America. The Additive homomorphism of this algorithm refers the same key for encryption but uses the random number of k in Eg() which makes AHEE able to resist plaintext attack. The AHEE is the subset of the fully homomorphism. Its applications are secure multi-party computation, electronic voting and mobile cipher.

**Transfer phase**

1. client encrypts the data using AHEE algorithm.
2. 2 keys are generated, those are a and b. These 2 keys are known only to the client.
3. client again encrypts the data using the algorithm used by the server/third party and sends it to the cloud storage.

Since the data was encrypted by client using AHEE algorithm previously, the encrypted data will be stored in the database. If any intruder hacks the data, he cannot access it since it is encrypted.

**Retrieval phase**

1. If the user needs the particular data, he/she downloads the data from the cloud.
2. The client decrypts the data using the algorithm used by the server.
3. The client again decrypts the data using AHEE algorithm. Here the keys a and b are used which are known only to the client

**Result**

Successful outcome of our project:

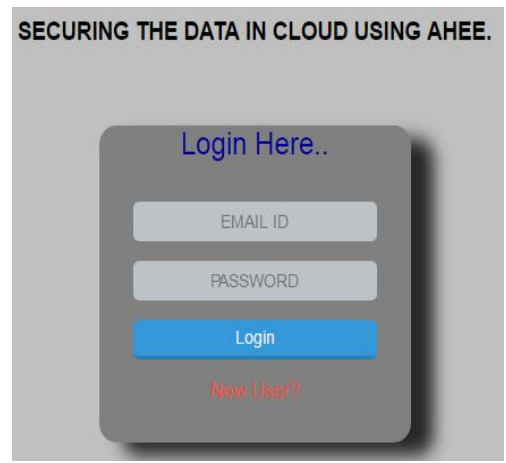

**Figure 4:** Login Page

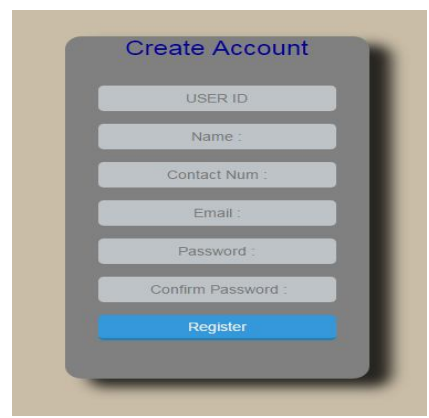

**Figure 5:** Registration Page

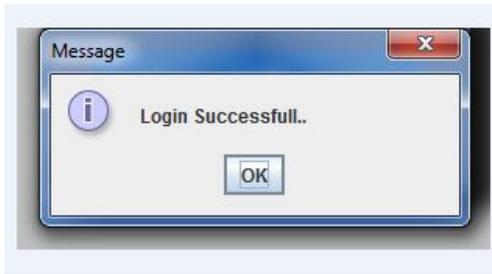

**Figure 6:** Successful login

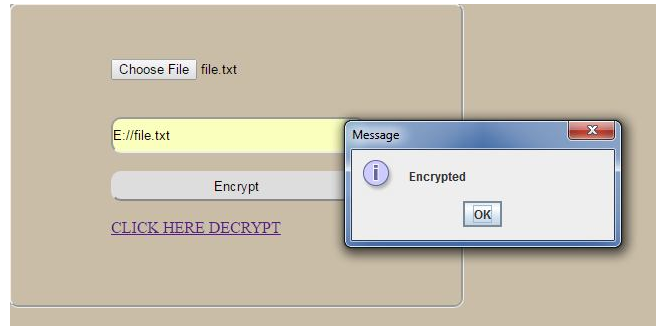

**Figure 10:** Encryption Successful

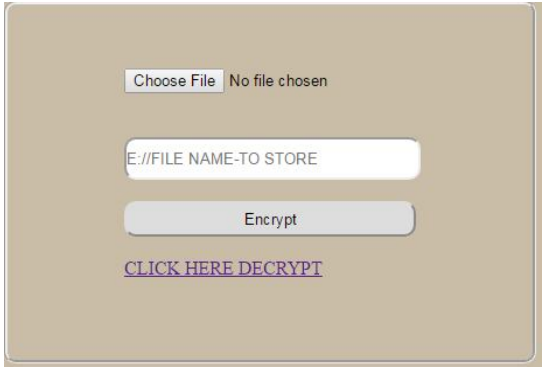

**Figure 7:** Encryption page

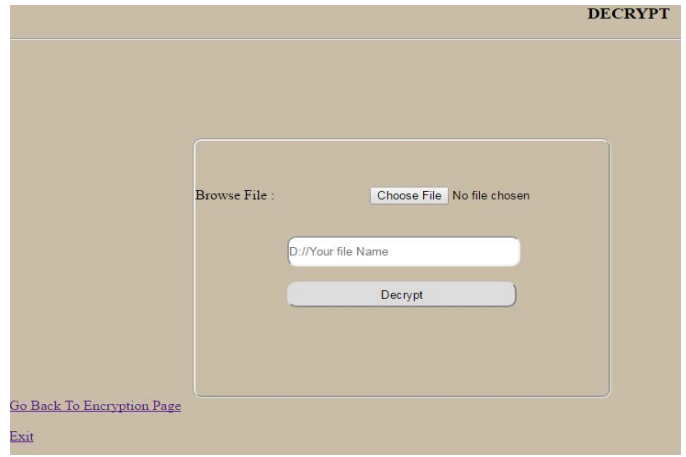

**Figure 11:** Decryption Page

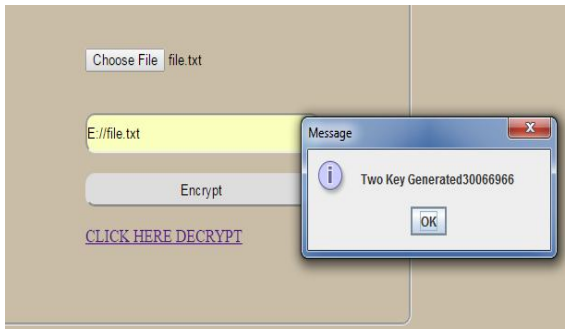

**Figure 8:** First Key Generated

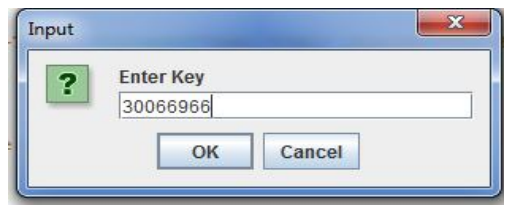

**Figure 12:** Entering the first key

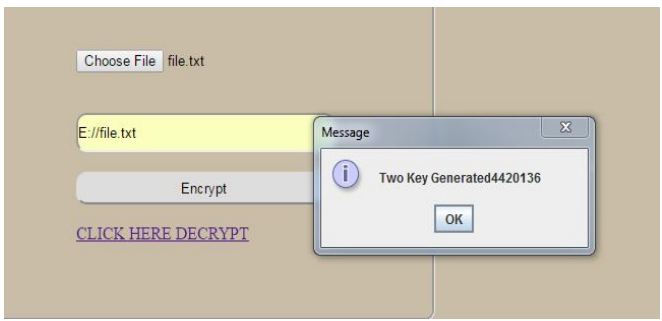

**Figure 9:** Second Key Generated

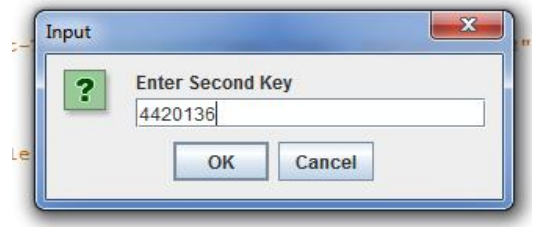

**Figure 13**: Entering the first key

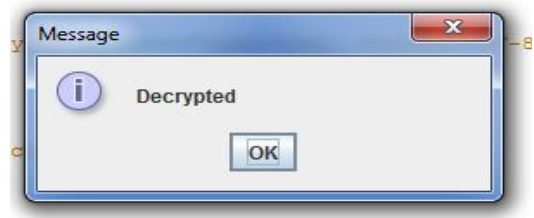

**Figure 14:** Decryption successful

## 5. CONCLUSION AND FUTURE ENHANCEMENT

Cloud Computing is becoming the next stage platform in advancement of internet. It provides the customer an effective and efficient way to store the data in the cloud with different capabilities and applications. The data in the cloud is stored by the service provider. A service provider is capable and having a technique to protect their client data to ensure security and to stop the data from disclosure the unauthorized users. But, we cannot fully depend on service providers to secure our data. Hence, we have used Algebra Homomorphic Encryption scheme based on updated Elgamal to provide security to the data on the cloud. With more advancements in cloud computing, one can-not overlook the issue of Security. There are various mechanism to provide cloud security, and thus improve the QoS related to cloud computing. We in this paper have tried to show the working of a standard AHEE algorithm in securing the data before it can be put out in the cloud.

As the algorithm is light weight protocol, it can be used for wireless sensor networks, smart grid environment etc. which are dependent on the cloud computing for sharing the vital information.